\documentstyle[prb,aps,preprint,tighten]{revtex}

\begin{document}
\draft

\title{Exact Boundary Critical Exponents and Tunneling Effect in Integrable
Models for Quantum Wires}
\author{Y. Wang$^{1}$\cite{*}  J. Voit$^{1,2}$  and Fu-Cho Pu$^3$}
\address{1. Physikalisches Institut, Universit\"{a}t Bayreuth, 
D-95440 Bayreuth, Germany\\
2. Bayreuther Institut f\"{u}r Makromolek\"{u}lforschung (BIMF), 
Universit\"{a}t Bayreuth, \\
D-95440 Bayreuth, Germany\\
3. Institute of Physics, Chinese Academy of Sciences, Beijing 100080, P. R. 
China}%

\maketitle
\begin{abstract}%
Using the principles of the conformal quantum field theory and the finite size
corrections of the energy of the ground and various excited states, 
we calculate the boundary critical exponents of single- and multicomponent
Bethe ansatz soluble models. The boundary critical exponents are given in 
terms 
of the dressed charge matrix which has the same form as that of systems with
periodic boundary conditions and is uniquely determined by the Bethe ansatz 
equations. A Luttinger liquid with open boundaries is the effective low-energy
theory of these models.
As applications of the theory, the Friedel oscillations due to the boundaries 
and the tunneling conductance through a barrier are also calculated. 
The tunneling conductance is determined by a nonuniversal boundary exponent 
which governs its power law dependence on temperature and frequency.
\end{abstract}
\narrowtext

\newpage

\section{Introduction}
Phase transitions take place in a different way on surfaces and in the bulk of
a sample \cite{11}. The exponents describing critical phenomena at surfaces 
differ from those of the bulk, and one may observe new phenomena due to 
anisotropy and the breaking of translational invariance caused by the 
boundary, like oscillations
in correlation functions which, in the bulk, are monotonous, coordinate 
dependences, and
in particular Friedel oscillations, in local quantities, etc. There are many 
approximate
methods for calculating the critical properties of bulk and surface phenomena.

Systems at a bulk critical point are not only scale invariant but also 
conformally invariant, a 
consequence of the combined rotational and scale invariance \cite{1,2}. 
In two space
dimensions or one space and one time dimension [thus including 
one-dimensional (1D) quantum
systems],
the constraints imposed by conformal invariance are much stronger than 
in higher dimensions
because the conformal group is infinite-dimensional, and these constraints 
strongly simplify the calculation of correlation functions. The conformal
field theory is parametrized by a unique constant -- the conformal anomaly 
or the central charge $c$
of the corresponding Virasoro algebra \cite{3}. All critical exponents of 
theories with
$c < 1$ are universal (independent of the interaction) and can be calculated 
exactly \cite{4}.
However, for $c \geq 1$, the conformal dimensions continuously
depend on the coupling of the fields and there is no general theory to deduce 
the critical exponents exactly. Conformal field theory only predicts that the
central charge and the conformal dimensions can be derived from the finite 
size corrections to the energy and momentum spectra \cite{5,6,7}. For 
$ c \geq 1$, it does not determine their actual numerical values. Still,
this constitutes a powerful method to calculate the critical exponents
of integrable 1D quantum models because their energy and momentum spectra are 
known exactly, and this strategy has been applied successfully for bulk
properties \cite{9,10}.

An entirely parallel and equivalent development has taken place in the theory 
of correlated
fermions in 1D, using bosonization techniques and running under the name of 
``Luttinger liquid''
\cite{Haldane,myreview}. The main focus here was on non-Fermi liquid 
properties of strongly
correlated fermions, an  exciting topic of current research. Fermi liquid 
theory fails in 1D,
and the Luttinger liquid provides the universal low-energy theory for 
gapless 1D quantum systems.
Its salient properties are: (i) absence of fermionic quasi-particle 
excitations, (ii) anomalous
dimensions of various operators leading to non-universal power-law decay of 
correlation 
functions, and (iii) charge-spin separation. In terms of critical phenomena, 
the system is at
a $T=0$-quantum critical point, is conformally invariant, and the central 
charge of the
associated Virasoro algebra(s) is unity. The exponents of the correlation 
functions (critical
exponents) are related to each other by scaling relations and depend on a 
single renormalized
coupling constant $K$ per degree of freedom, playing the role of the Landau 
parameters familiar
from Fermi liquid theory. For the Luttinger liquid, there are constitutive 
relations between
three velocities characterizing the low-energy sector of the spectrum of the 
Hamiltonian
which determine the renormalized coupling constant and thus the critical 
exponents. For
integrable models, the velocities and coupling constants have been determined 
from Bethe
Ansatz (or other) solutions \cite{map}. In some cases, conformal invariance 
has been used 
explicitly to determine the critical exponents \cite{19}.
For non-integrable models, they can be obtained reliably
by exact diagonalization of the Hamiltonian. 

Recently, the problems of the one-dimensional systems with open boundaries 
have drawn
much attention. There are several issues of both experimental and theoretical 
relevance. 
(i) As in higher-dimensional systems,
the boundary critical exponents are expected to be different from the bulk 
ones \cite{11}.
(ii) Experimentally studied systems are finite, and the progress in 
microfabrication of 
semiconductor structures has provided us with quantum wires so small that
the boundary effects could become relevant \cite{12}. (iii) Here but also 
in the much bigger
samples of 
quasi-1D metals, impurities can be relevant perturbations and effectively 
cut the systems
into isolated strands of finite length. Specifically, a renormalization 
group analysis 
has shown that, for effectively repulsive interactions, 
the scattering potential due to an isolated impurity scales to 
infinitity in the low energy 
regime and
thus the problem is equivalent to an open boundary problem \cite{13}. 
An attempt to
test experimentally the predictions of this theory has been 
published recently \cite{mill}.
Such effects have also been
invoked in the interpretation of electron spin diffusion \cite{dormann} and
photoemission \cite{phot,johan} experiments on quasi-1D organic conductors. 
(iv) Finally, some numerical methods such as the density-matrix 
renormalization group \cite{white}, rely on 
the use of open boundaries and their results could, in principle, 
be affected by the chain ends.

Both conformal field theory \cite{14} and, very recently, the theory of 
Luttinger liquids
\cite{12}, have been extended to systems with boundaries. What is 
missing to date, however,
is an exact derivation of the boundary critical exponents 
from the Bethe ansatz solution of 
integrable quantum systems bridging the gap between microscopic 
(often lattice) models containing both high and low-energy physics,  and the 
more effective theories for the low energy properties. This gap has been 
bridged successfully for the periodic systems \cite{myreview,map,19}. 
It is the purpose of this paper to present such an
exact derivation of boundary critical exponents. Moreover, 
in the course of the study, we 
shall see that the same strategy can be applied to determine the 
exponents of nonintegrable
systems by exact numerical diagonalization provided they satisfy the basic 
assumption of conformal invariance.

In this paper, we apply the method of the conformal field theory to 
the Bethe ansatz
soluble models with open boundaries. The layout of the present 
paper is the following.
In the following Section we briefly summarize some important results 
of boundary conformal
field theory and of Luttinger liquid theory on bounded systems, in 
order to make the presentation
self-contained and provide the basic tools. These are essentially 
the finite size corrections
of the energy spectrum. In Section III we give a detailed calculation 
of the boundary 
critical exponents of two paradigmatic single-component Bethe ansatz 
soluble models 
($\delta$-potential Bose gas and the antiferromagnetic 
Heisenberg chain). Section IV 
digresses to two important physical applications: the 
Friedel oscillation of the density distribution around the boundary 
(impurity) and
the tunneling conductance through a barrier in a quantum wire. In
Section V we generalize the result to the multicomponent case with the 
Hubbard model as an 
example. The summary in Section VI attempts to provide a broader perspective 
on our results.

\section{Conformal field theory and Luttinger liquids in bounded systems}
Systems with open boundaries 
\begin{equation}
\label{obc}
\psi ( x=0 ) = \psi (x=L) = 0
\end{equation}
(where $\psi$ is the wavefunction)
are no longer space translational invariant but the time 
translational invariance is preserved. 
A two-point correlation function of a (primary) field at criticality then 
takes the general form
\cite{14} 
\begin{eqnarray}
\label{twoptfct}
G_b(\tau, x_1, x_2) = (x_1x_2)^{-d}\Phi\{[v^2\tau^2+x_1^2+x_2^2]/x_1x_2\} \; .
\end{eqnarray}
Eq.~(\ref{twoptfct}) applies to a 1D quantum system, and $\tau$ represents 
the imaginary time 
and $x_{1,2}$ the spatial coordinates; $d$ is the conformal dimension
of the primary field in the bulk; $v$ is the Fermi velocity. When $x_1$ and 
$x_2$ are near the 
boundary, and $\tau \to \infty$, $G_b$ must behave as
\begin{eqnarray}
\label{asympt}
G_b(\tau, x_1, x_2) \sim \frac 1{\tau^{2x_b}} \;.
\end{eqnarray}
$2x_b$ is the boundary critical exponent. Eq.~(\ref{asympt}) directly 
implies that 
$\lim_{y\to\infty}\Phi(y)\sim y^{-x_b}$. On the other hand, Cardy also 
showed that
the $n$-point correlation function in a half plane (with one open boundary 
at $x=0$) is 
identical to the $2n$-point correlation function in the whole plane, 
provided only the 
$z$-dependent
part is taken into account in the latter. In this way, the two-point 
correlation function
can also be represented as
\begin{eqnarray}
\label{twoptalt}
G_b(z_1,z_2,\bar{z}_1,\bar{z}_2)=
[(z_1-\bar{z}_1)(z_2-\bar{z}_2)/(z_1-z_2)(\bar{z}_1-\bar{z}_2)
(z_1-\bar{z}_2)(\bar{z}_1-z_2)]^{-2d}F_b(y) \; ,
\end{eqnarray}
where $F_b(y)$ is an unknown scaling function. Here, we have switched 
to a notation in terms of
complex variables $z_j=v\tau_j+ix_j$, $\bar{z}_j=v\tau_j-ix_j$ and 
$y$ is given by
$y=(z_1-z_2)(\bar{z}_1-\bar{z}_2)/(z_1-\bar{z}_1)(z_2-\bar{z}_2)$.
For $y\to\infty$, $F_b(y) \to y^{-\alpha}$. Direct comparison to 
Eq.~({\ref{twoptfct}) gives
\begin{eqnarray}
\label{scaling}
x_b = 4d + \alpha \;.
\end{eqnarray}
In the following text, we shall use Eq.~(\ref{twoptfct}) and 
Eq.~(\ref{twoptalt}) alternatively.

The conformal dimensions or critical exponents can be calculated 
from the finite size corrections
of the energy spectra. To see this, consider the transformation
\begin{eqnarray}
\label{transf}
\zeta=\frac L\pi\ln z, \; \; \; \; \; \; \; \; \; \; 
\bar{\zeta}=\frac L\pi\ln \bar{z} \;,
\end{eqnarray}
applied to the upper half-plane $x \geq 0$ only. Such a 
conformal transformation maps the system 
from the semi-infinite plane onto a strip of width $L$ 
with open boundary conditions  \cite{5}. 
From the general transformation properties of the correlation functions of a 
(primary) conformal field $\phi(z,\bar{z})$
\begin{eqnarray}
\label{gentf}
\langle \phi(z_1,\bar{z}_1)\phi(z_2,\bar{z}_2) \rangle =
(\frac{\partial\zeta_1}{\partial z_1})^\Delta 
 (\frac{\partial\bar{\zeta}_1}{\partial \bar{z}_1})^{\bar{\Delta}}
(\frac{\partial\zeta_2}
 {\partial z_2})^\Delta 
 (\frac{\partial\bar{\zeta}_2}{\partial \bar{z}_2})^{\bar{\Delta}}
\langle \phi(\zeta_1,\bar{\zeta}_1)\phi(\zeta_2,\bar{\zeta}_2) \rangle \; ,
\end{eqnarray}
where $\Delta+\bar{\Delta}=d$ and $\phi$ and $\Delta-\bar{\Delta}=s$ 
give, respectively, the conformal dimension
and spin of the field, one deduces for the correlation function on the strip
\begin{eqnarray}
\langle \phi(z_1,\bar{z}_1)\phi(z_2,\bar{z}_2) \rangle_L & = & 
[\frac{\frac{\pi}{4L} \sinh\frac{\pi(z_1-\bar{z}_1)}{2L}
\sinh\frac{\pi(z_2-\bar{z}_2)}{2L}}{\sinh\frac{\pi(z_1-z_2)}{2L}
\sinh\frac{\pi(\bar{z}_1-\bar{z}_2)}{2L}\sinh\frac{\pi(z_1-\bar{z}_2)}{2L}
\sinh\frac{\pi(\bar{z}_1-z_2)}{2L}}]^{2d} \nonumber \\
\label{stripfct}
& \times & F_b(\frac{\sinh\frac{\pi(z_1-\bar{z}_1)}{2L}
\sinh\frac{\pi(z_2-\bar{z}_2)}{2L}}{\sinh\frac{\pi(z_1-z_2)}{2L}
\sinh\frac{\pi(\bar{z}_1-
\bar{z}_2)}{2L}}) \; .
\end{eqnarray}
There is also a spectral representation of the correlation 
function on the strip
\begin{eqnarray}
\label{stripspect}
 \langle \phi(z_1,\bar{z}_1)\phi(z_2,\bar{z}_2) \rangle _L=
\sum_n \langle \phi(x_1,0)|n \rangle  \langle n|\phi(x_2,0) \rangle
\exp[-(\tau_1-\tau_2)(E_L^n-E_L^0)] \; ,
\end{eqnarray}
where $E_{b,L}^0$ is the energy of the ground state and $E_{b,L}^n$ are the 
energies of the excited states; $|n \rangle $ are the exact eigenstates of 
the Hamiltonian under consideration which form a
complete set. Suppose $E_L^1$ (the energy of the first excited state with the
form factor $\langle \phi|1 \rangle \neq0$) 
takes the form ($L\to\infty$),
\begin{eqnarray}
\label{finsiz}
E_L^1-E_L^0=\frac{\pi v}Lx_b+o(\frac1{L}).
\end{eqnarray}
From the asymptotic form of the correlation functions Eq.~(\ref{asympt}) 
and comparing
Eqs.~(\ref{stripfct}) and (\ref{stripspect}) we get Eq.~(\ref{scaling}). 
Thus the finite size asymptotics of the low lying levels
determines the boundary critical exponents.

In general, the correlation function can oscillate, so that its
asymptotics is not conformally invariant. In that case, however, one can
decompose the field $\phi(z, \bar{z})$ into a sum of conformal fields
$\phi_n(z,\bar{z})$
which then determine the power-law asymptotics \cite{9}. Because of the 
reflection symmetry of the open boundary systems, the field $\phi$ must have 
definite parity. We can expand odd and even parity fields as
\begin{eqnarray}
\phi(z,\bar{z})=\sum_n\phi_n(z,\bar{z})\sin(nk_Fx) \; ,
\hspace{0.5cm}
\phi(z,\bar{z})=\sum_n\phi_n(z,\bar{z})\cos(nk_Fx) \; ,
\end{eqnarray}
respectively.
$n$ is an odd (even) integer for odd (even) parity fields.

For systems with periodic boundary conditions, the Luttinger liquid 
phenomenology \cite{Haldane,myreview} provides a framework completely
equivalent to conformal field theory but closer to the language of 
conventional solid
state physics. It is based on the exactly solvable Luttinger model, and
all physical properties can be described in terms of two parameters per
degree of freedom ($\nu = \rho, \sigma$ for charge and spin), a renormalized 
sound velocity
$v_{\nu}$ and an effective coupling constant $K_{\nu}$ which determines
the decay of all correlation functions and thus the critical exponents. 
These parameters can be determined from the energies of the low-lying
excited states of the Hamiltonian \cite{map}.

This picture has been extended recently to systems with open boundaries
\cite{12}. Due to the boundary conditions (\ref{obc}), the right- and 
left-moving fermions commonly used in the Luttinger model are not independent, 
and a single species moving, say, to the right [$\Psi_{+,s}(x)$] 
is sufficient, and
it is periodic on a length $2L$. The Fermi surface reduces to a single 
point $+k_F$ but the wavevectors $k = m \pi
/ L >0$ are quantized with twice the density of the periodic system. 
We then can rewrite the fermionic Hamiltonian
\begin{equation}
\label{lbferm}
H_0 = - i v_F \sum_s \int_{-L}^L \! dx \Psi_{+,s}^{\dag}(x) \partial_x
\Psi_{+,s}(x)
\end{equation}
in an equivalent form involving the bosonic density fluctuations 
(particle-hole 
excitations) $\rho_{+,s}(x)$ and ``charge excitations'' $\Delta N_s$
corresponding to the addition of particles of spin $s$ to the reference
Fermi sea (i.e. $\Delta k_{Fs} = \Delta N_s \pi / L$) 
\begin{equation}
\label{lbbos}
H_0 = \pi v_F \sum_s \int_{-L}^L \! dx : \rho_{+,s}(x) \rho_{+,s}(x) :
+ \frac{\pi v_F }{2L} \sum_s \left( \Delta N_s \right)^2 \; .
\end{equation}
The Fourier transform $\rho_{+,s}(p)$ of the density operators do not 
contain the
$p=0$-component which is represented explicitly by $\Delta N_s = \sum_k 
( c^{\dag}_{+,s,k}
c_{+,s,k} - \langle c^{\dag}_{+,s,k}
c_{+,s,k} \rangle_0 )$ where $\langle ... \rangle_0$ denotes the 
(infinite) expectation value in the
reference Fermi sea given by $k_F^0$. Unlike the periodic case, 
``current excitations''
describing the difference of right- and left-moving fermion numbers, 
cannot be defined in the bounded system.
The Hamiltonian including forward scattering can 
then be diagonalized by a Bogoliubov tranformation as in the 
periodic case, defining the
renormalized velocities $v_{\nu}$ of the bosonic charge and spin 
density fluctuations
$\rho_+(p) [\sigma_+(p)] = [\rho_{+,\uparrow}(p) \pm 
\rho_{+,\downarrow}(p)]/\sqrt{2}$,
and coupling constants $K_{\nu}$. The renormalized velocity of the 
charge excitations 
$\Delta N_{\rho (\sigma)} = \Delta N_{\uparrow} \pm \Delta N_{\downarrow}$ is
given by $v_{\nu} / K_{\nu}$. 
The bosonization of this model is completed by an
explicit representation of the Fermi operator $\Psi_{+,s}(x)$ in terms 
of the bosons
$\rho_{+,s}(p)$ \cite{12} which allows to calculate all correlation 
functions of this model
in terms of the $v_{\nu}$ and $K_{\nu}$ and thus defines its critical 
exponents.

$v_{\nu}$ and $K_{\nu}$ can now be found along the same 
lines as in the periodic systems
\cite{map}:
(i) $v_{\nu}$ can be computed directly from the spectrum 
of low-lying excitations; (ii) 
to get $K_{\rho}$, one calculates the compressibility $\kappa$
\begin{equation}
\label{compr}
\frac{1}{\kappa} = \frac{1}{L} \frac{\partial^2 E_0(n)}{\partial n^2} = 
\frac{2 K_{\rho}}{\pi v_{\rho}} \; .
\end{equation}
The first equality gives the definition as the second 
derivative of the ground state energy
with respect to particle density $n = N/L$ and can be computed in the 
integrable model,
and the second equality gives the Luttinger liquid expression which can 
be solved for 
$K_{\rho}$. $K_{\sigma}$ is required to be unity by spin-rotation invariance,
but can be calculated in the same way in
more general cases from the susceptibility. Our determination of boundary 
critical exponents below can also be viewed as exploring this strategy.

\section{Bethe ansatz soluble models}
We now compute the exact boundary critical exponents for Bethe ansatz 
soluble models with open
boundaries. We first consider single-component models before turning to 
multicomponent systems.
In both cases, the critical exponents can be calculated explicitly. 
Two typical single-component models are the one dimensional 
$\delta$-potential Bose gas model and the Heisenberg
antiferromagnetic chain. Their Hamiltonians are
\begin{eqnarray}
H_{BG}=\int_0^L \left(\partial_x\Psi^\dagger \partial_x\Psi + c\Psi^\dagger
\Psi^\dagger\Psi\Psi-h\Psi^\dagger\Psi \right ) dx, {~~~}c>0,{~~~}h>0,
\end{eqnarray}
\begin{eqnarray}
\label{hxxz}
H_{XXZ}=\sum_{j=1}^{L-1}(\sigma_j^x\sigma_{j+1}^x+\sigma_j^y\sigma^y_{j+1}+
\cos2\eta \sigma_j^z\sigma_{j+1}^z-\frac12h\sigma_j^z),\\
0<2\eta<\pi,{~~~}0<h<4(1-\cos2\eta),\nonumber
\end{eqnarray}
where $h$ is the chemical potential for the Bose gas model and the magnetic 
field for the Heisenberg chain. The anisotropy of the exchange integrals 
for the Heisenberg model is $J_z = \cos(2 \eta)$; critical behavior is 
obtained only for 
easy-plane-type anisotropy $| J_z | \leq 1$, and our definition of $J_z$ 
restricts us
to this range. $\eta$ plays the role of a coupling constant, as is $c$ 
for the Bose gas.
${\sigma^{x,y,z}}$ are the
Pauli matrices. The open boundary conditions (\ref{obc}), for the Bose 
gas, translate into
$\Psi(0) = \Psi(L) = 0$ in terms of the boson operators $\Psi(x)$, while 
they are contained
in our representation (\ref{hxxz}) of the Heisenberg chain because the 
sites $1$ and $N$ only
couple to a single neighbor.

These models are solved by means of the Bethe ansatz \cite{15}. The 
$N$-particle wave function
is parametrized by $N$ numbers $\lambda_j$ which satisfy the equations
\begin{eqnarray}
\label{bethans}
2Lp_0(\lambda_j)=2\pi I_j- 2 \varphi(\lambda_j) -
\sum_{l\neq j}[\Phi(\lambda_j-\lambda_l)+\Phi(\lambda_j+\lambda_l)],
\end{eqnarray}
where $p_0$ is the bare momentum and $\Phi$ is the bare scattering phase:
\begin{eqnarray}
p_0^{BG}(\lambda)=\lambda,{~~~~~}
p_0^{XXZ}(\lambda)=i\ln(\frac{\cosh(\lambda-i\eta)}{\cosh(\lambda+i\eta)}),
\nonumber\\
\Phi^{BG}(\lambda)=-2\arctan\frac\lambda c,\\
\Phi^{XXZ}(\lambda)=-\pi+i\ln(\frac{\sinh(\lambda+2i\eta)}{\sinh(\lambda-
2i\eta)}).
\nonumber
\end{eqnarray}
The numbers $I_j$ are positive integers, and the parity effects known 
from periodic systems
are absent in models with open boundaries. The bare energy of each particle is
\begin{eqnarray}
\label{epsi}
\epsilon_0^{BG}(\lambda)=\lambda^2-h,{~~~} \epsilon_0^{XXZ}(\lambda)=
h-2\sinh^22\eta/\cosh(\lambda+i\eta)\cosh(\lambda-i\eta) \; .
\end{eqnarray}
The phase $\varphi(\lambda_j) = 0$ for the Bose gas and $\varphi(\lambda_j) =
p_0(\lambda_j) + \Phi(2 \lambda_j)$ for the XXZ-chain.
The eigenvalue of the Hamiltonian is equal to the sum of the bare energies of 
the particles
\begin{eqnarray}
E_L=\sum_{j=1}^N\epsilon_0(\lambda_j).
\end{eqnarray}
Carefully checking the wave functions with the boundary conditions 
Eq.~(\ref{obc}) we find that $\{\pm\lambda\}$ correspond to the same 
state. This is not surprising because of the 
reflection symmetry of the system, and corresponds to standing-wave-like 
solutions. The Bethe 
ansatz equation (\ref{bethans}) thus only allows solutions with
$\lambda_j\neq\pm\lambda_l$ for $j\neq l$ so that all
$\lambda_j\geq 0$. The system therefore has only one Fermi point $k_F$. This 
is very
different from the case of the cyclic systems. A similar feature, however, 
occurs in Luttinger
liquids in bounded systems (Section II). The appearence of a single Fermi 
point identifies 
the system as chiral, and 1D quantum systems with boundaries 
therefore appear to be special
cases of ``chiral Luttinger liquids'', a notion that has appeared 
previously in the 
superficially quite unrelated area of edge states in the fractional 
quantum Hall effect
\cite{wenrev}.

With open boundaries, the total
momentum is no longer a good quantum number.
However, the quantity
\begin{eqnarray}
P=\frac \pi L\sum_{j=1}^N|I_j|
\end{eqnarray}
is conserved. We shall call it the ``momentum'' of the models. 
In the ground state, $I_j$
takes consecutive integer values from 1 to $N$ ($I_j$=0 is not allowed). 
We can then define a dressed momentum
\begin{eqnarray}
\label{dressmom}
p_b(\lambda_j)=p_0(\lambda_j)+\frac 1{2L}\sum_{l\neq j}[\Phi(\lambda_j-
\lambda_l)+ \Phi(\lambda_j+\lambda_l)] + \frac{\varphi(\lambda_j)}{L}= 
\frac{\pi I_j}{L}\; .
\end{eqnarray}
The Fermi momentum $k_F=\pi N/L=\pi n$ has the same value as in the 
periodic system. 

In the thermodynamic limit $L \to \infty$, $N \to \infty$ keeping $n$ finite, 
the ground state solves the following integral equations
\begin{eqnarray}
\label{rhob}
\rho_b(\lambda)=\frac{p_0'(\lambda)}\pi+\int_0^\Lambda K_b(\lambda,\mu)
\rho_b(\mu)d\mu \; , \\
\label{epsint}
\epsilon_b(\lambda)=\epsilon_0(\lambda)+\int_0^\Lambda K_b(\lambda,\mu)
\epsilon_b(\mu)d\mu,\\
K_b(\lambda,\mu)=\frac 1{2\pi}\frac \partial{\partial\lambda}
[\Phi(\lambda-\mu)+ \Phi(\lambda+\mu)] \; .
\nonumber
\end{eqnarray}
The cutoff parameter $\Lambda$ is defined by the requirement 
$\epsilon_b(\Lambda)=0$; 
$\rho_b(\lambda)$ is
the density of $\lambda$ per unit length. Eq.~(\ref{dressmom}) becomes
\begin{eqnarray}
\label{drmomcon}
p_b(\lambda)=p_0(\lambda)+\frac12\int_0^\Lambda[\Phi(\lambda-\mu)+
\Phi(\lambda+\mu)]\rho_b(\mu)d\mu \; .
\end{eqnarray}
$p_b'(\lambda)=\pi\rho_b(\lambda)$ follows from a combination of (\ref{rhob}) 
and 
(\ref{drmomcon}), and $p_b(\Lambda)=k_F$ ensures that the Fermi 
surface is the same as in the finite system with the same electron density. 

We now compare equivalent quantities of the bounded and periodic systems 
and use the same symbols but without the subscript ``$b$" for the latter. 
Directly comparing 
Eqs.~(\ref{rhob})-(\ref{drmomcon})
with their periodic equivalents \cite{9} we find
\begin{eqnarray}
\rho_b(\lambda)=2\rho(\lambda) \; ,\nonumber\\
\label{compare}
\epsilon_b(\lambda)=\epsilon(\lambda) \; ,\\
p_b(\lambda)=p(\lambda) \; , \nonumber
\end{eqnarray}
if $n=n_b$ (thus $\Lambda_b=\Lambda$). There is no restriction on 
sign$(\lambda)$ for the
periodic systems but $\lambda > 0$ for the bounded ones.
In the vicinity of the Fermi surface, the excitation energy is
\begin{eqnarray}
e(\lambda)=v|p_b(\lambda)-k_F| \; ,
\end{eqnarray}
where the Fermi velocity is given by
\begin{eqnarray}
v=\frac{\epsilon'(\Lambda)}{p'(\Lambda)}=\frac{\epsilon'_b(\Lambda)}{\pi
\rho_b(\Lambda)}.
\end{eqnarray}
It takes the same value of that of the system with the periodic boundary 
condition
as it must be for the Fermi velocity can be determined by the leading 
term of the free energy 
which should not depend on the boundary conditions in the thermodynamic limit.

Unlike the cyclic systems, the systems with open boundaries have only two kinds
of elementary excitations: (i) Particle-hole (sound-like) excitations 
at the Fermi point $k_F$. Their finite size corrections give 
the boundary critical exponent
of the current-current correlation function for the Bose gas model and or 
the boundary
critical exponent of the $S_z$-component of the spin correlation function 
for the Heisenberg spin
chain. (ii) The change of the free energy induced by the variation of the 
particle number
(termed ``charge excitations'' above). Its finite size correction gives the 
boundary critical exponent of the single particle correlation function for the
Bose gas model and the critical exponent of the transverse spin-spin 
correlation function
of the Heisenberg chain. These features are reproduced precisely in the 
Luttinger liquid
theory of systems with open boundary conditions.

We consider first the particle-hole excitations. To construct the lowest 
excitation
state, we must put $I_N\to I_N+1$ in (\ref{bethans}) keeping the other 
$I_j$ unchanged. The
change of the momentum is thus
\begin{eqnarray}
\Delta P=\frac\pi L,
\end{eqnarray}
and the excitation energy is
\begin{eqnarray}
\Delta E_{b,L}=\frac{\pi v x_b^\parallel}L, {~~~~~~~~~} x_b^\parallel=1.
\end{eqnarray}
To obtain the second kind of excitations (charge excitations), we define the 
dressed charge 
function as
\begin{eqnarray}
\label{dressed}
Z_b(\lambda)=1+\int_0^\Lambda K(\lambda,\mu)Z_b(\mu)d\mu.
\end{eqnarray}
Obviously, $Z_b(\lambda)=Z(\lambda)$ for $\lambda > 0$. The change of the 
free energy by
$\Delta N$ additional particles is \cite{9}
\begin{eqnarray}
\Delta E_{b,L}=L[f_0(n+\frac{\Delta N}L)-\Delta N\frac h L-f_0(n)]=
\frac{(\Delta N)^2}{2L}
\frac{\partial h}{\partial n},
\end{eqnarray}
where $f_0$ is thefree energy density of the ground state. This gives 
\begin{eqnarray}
\Delta E_{b,L}=\frac{\pi v(\Delta N)^2}{2LZ^2(\Lambda)}, 
{~~~~}x_b^\perp=\frac{(\Delta N)^2}{2Z^2(\Lambda)}.
\end{eqnarray}
For the general case, suppose a conformal field $\phi$ induce the 
momentum shift
relative to the ground state as $\pi\Delta I/L$ and the change of 
the particle number
$\Delta N$. The energy change relative to the ground state is then
\begin{eqnarray}
\Delta E_{b,L}=\frac{\pi v}L[\Delta I+\frac{(\Delta N)^2}{2Z^2(\Lambda)}]+
o(\frac 1{L}).
\end{eqnarray}
The boundary critical exponent takes the form
\begin{eqnarray}
2x_b(\Delta I,\Delta N)=2 \Delta I+\frac{(\Delta N)^2}{Z^2(\Lambda)},
\end{eqnarray}
where $\Delta I$ and $\Delta N$ are non-negative integers. 
The above relation was
suggested by Alcaraz \em et  al. \rm \cite{alca} for integrable spin 
chains from numerical simulations.

For example, we consider the  Heisenberg chain in a zero magnetic field.
In this case \cite{16}, $\Lambda=\infty$ and $2Z^2(\infty)=\pi/2\eta$. 
For $n_1$, $n_2$
near the boundary and $t \to \infty$, the correlation functions take the 
following asymptotic forms 
\begin{eqnarray}
\langle \sigma_{n_1}^z(t)\sigma_{n_2}^z(0) \rangle 
\sim\frac{1}{t^{\eta_\parallel}} \; ,
\nonumber\\
\Delta I=1,{~~~} \Delta N =0, {~~~}\eta_\parallel=2x_b^\parallel=2 \; ;
\end{eqnarray}
\begin{eqnarray}
\langle \sigma_{n_1}^-(t)\sigma_{n_2}^+(0) \rangle 
\sim\frac1{t^{\eta_\perp}} \; , \nonumber\\
\Delta I =0,{~~~} \Delta N=1,{~~~}\eta_\perp=2x_b^\perp=\frac{4\eta}\pi \; .
\end{eqnarray}
On the other hand, for $t=0$, $n_1=1$ and $n_2=n \gg 1$, from 
Eq.~(\ref{twoptfct}) we have
\begin{eqnarray}
\langle \sigma_1^z(0)\sigma_n^z(0) \rangle
\sim\frac{(-1)^{n-1}}{n^{\theta_\parallel}} \; ,
\nonumber\\
\theta_\parallel=1+\frac\pi{4\eta}\; ,{~~~}{\rm for} {~~~}2\eta>\frac\pi2\\
\theta_\parallel=2 \; ,{~~~} {\rm for} {~~~}2\eta\leq\frac\pi2\nonumber
\end{eqnarray}
and
\begin{eqnarray}
\langle \sigma^-_1(0)\sigma^+_n(0) \rangle 
\sim\frac{(-1)^{n-1}}{n^{\theta_\perp}},\nonumber\\
\theta_\perp=\frac{3\eta}\pi \; .
\end{eqnarray}
Notice that above we have used the bulk conformal dimensions 
$d_\parallel=\frac\pi{4\eta}$
for $4\eta > \pi$, $d_\parallel = 1$ for $4\eta\leq \pi$ and 
$d_\perp=\eta/\pi$.
When $2\eta=\pi$, the coupling is isotropic and $\eta_\parallel=\eta_\perp=2$, 
$\theta_\parallel=\theta_\perp=\frac 32$ as they should be for the spin $SU(2)$
symmetry. The boundary critical exponents $\eta_{\|,\perp}$ measuring the 
decay of correlations with time, are two times larger than those of the bulk. 
On the other hand, there are new exponents $\theta_{\|,\perp}$ for the
decay of spatial correlations. In periodic systems, the two sets are identical
as a consequence of conformal invariance. In bounded systems, the breaking
of translational invariance along $x$ by the boundary conditions while 
maintaining it along $t$, generates a new set of critical exponents.

We now consider the finite size correction to the ground state energy
\begin{eqnarray}
E_{b,L}^0=\sum_{j=1}^N\epsilon_0 \left(\lambda[j / L] \right) \; .
\end{eqnarray}
Using the Euler-Maclaurin formula we have \cite{17}
\begin{eqnarray}
E_{b,L}^0=L\int_0^n \! \epsilon_0 \left(\lambda[x] \right) dx + f_b - 
\frac 1{24L}\frac{\partial\epsilon_0(x)}{\partial x}|_{x=n}+
\frac 1{24L}\frac{\partial\epsilon_0(x)}{\partial x}|_{x=0}+o(\frac 1{L}) 
\nonumber\\
=Lf_0(n)+f_b-\frac\pi{24L}[\frac{\epsilon'_b(\Lambda)}{\pi\rho_b(\Lambda)}-
\frac{\epsilon'_b(0)}{\pi\rho_b(0)}]+o(\frac 1{L}).
\end{eqnarray}
From Eqs.~(\ref{epsi}) and (\ref{epsint}) we know that $\epsilon'_b(0)=0$. 
Thus we have
\begin{eqnarray}
\label{ccharge}
E_{b,L}^0-Lf_0(n)\approx f_b-\frac{\pi v}{24L} \; .
\end{eqnarray}
Here $f_b$ is the boundary energy which was extensively studied 
\cite{15,hamer,20,21}.
Eq.~(\ref{ccharge}) agrees with the predictions of conformal field 
theory \cite{7}, and determines the central charge as $c=1$.

\section{friedel oscillation and tunneling conductance}

\subsection{The Friedel oscillation}
Since the systems under consideration are obviously not 
translationally invariant, the density 
distribution is no longer homogeneous.
Therfore, the ground state will exhibit Friedel oscillations. 
For a 1D free fermion
system of length $L$ with open boundaries (\ref{obc}), the single-particle 
wave functions take the form
\begin{eqnarray}
\Psi_m^L(x)=\sqrt{\frac2L}\sin\frac{ m \pi x}L,
\end{eqnarray}
with $m$ positive integers. For the ground state,
the density distribution in the box can be easily calculated as
\begin{eqnarray}
\langle n(x) \rangle =\sum_{m=1}^N|\Psi_m^L(x)|^2=\frac 2L\sum_{m=1}^N
\sin^2\frac{m \pi x}L\nonumber\\
\approx n -\frac{\sin(2k_Fx)}x,{~~~~~~~~~} {\rm for} {~~}x \ll L \; .
\end{eqnarray}
For the interacting systems, we expect the density distribution to 
have a similar form
\begin{eqnarray}
\label{friedel}
\langle n(x) \rangle_
b\approx  n -\frac{A\sin(2k_Fx-\phi)}{x^\gamma}, {~~~~~~}0 \ll x \ll L,
\end{eqnarray}
where $A$ and $\phi$ are two unknown constants and $\gamma$ is the exponent
dominating the decay. As pointed out in Ref. 22, 
the $n$-point correlation functions of the open boundary systems are 
directly related 
to the $2n$-point correlation functions of the periodic boundary systems. Thus
\begin{eqnarray}
\label{dens}
\langle n(x) \rangle_b= \langle n(z)n(\bar{z}) \rangle \; .
\end{eqnarray}
The right hand side of Eq.~(\ref{dens}) is to be understood in the sense 
that only the 
$z$-dependent part of the density-density
correlation function $ \langle n(z_1,\bar{z}_1)n(z_2,\bar{z}_2) \rangle$ 
is taken into account, and that we have
to set $z_2=\bar{z}_1$. In addition, the oscillating term only originates 
from the 
current-current correlation $ \langle n(z_1,\bar{z}_1)n(z_2,\bar{z}_2) 
\rangle_J$ (in another language
\cite{myreview}, this object is designed as the $2k_F$-charge density 
wave correlation function)
which has been calculated in Ref. 9 
as
\begin{eqnarray}
\label{rcdw}
\langle n(z_1,\bar{z}_1)n(z_2,\bar{z}_2) \rangle_J \approx
-\frac{Be^{i2k_F(x_1-x_2)}+
\bar{B}e^{i2k_F(x_2-x_1)}}{(z_1-z_2)^{Z^2(\Lambda)}(\bar{z}_1-
\bar{z}_2)^{Z^2(\Lambda)}},
\end{eqnarray}
where  $B$ is a constant. Choosing the $z$-dependent part in 
Eq.~(\ref{rcdw})
and putting $z_2=\bar{z}_1$, we obtain Eq.~(\ref{friedel}) with
\begin{eqnarray}
\gamma = Z^2(\Lambda)
\end{eqnarray}
This result qualitatively agrees with Ref. 15 
where the Friedel oscillation for the 
Luttinger model with open boundaries was calculated, and determines the 
exact value of $\gamma$ from the Bethe ansatz equations through the dressed 
charge which can be calcualted easily.

\subsection{The tunneling conductance}
The boundary critical exponents are very important to study the tunneling 
effect in 
quantum wires. A strong barrier cuts the chain into two half-chains 
which, in first order,
behave as two independent subsystems with an open boundary. 
For simplicity, we suppose them
to be spinless and their Hamiltonian then reads
\begin{eqnarray}
\label{spinless}
H=-\sum_{r=\pm}\sum_{j=1}^{N_r-1}
\left\{ \left( C_{r,j}^\dagger C_{r,j+1} + {\rm H.c.} \right) -
Un_{r,j}n_{r,j+1}
\right\}-h\sum_{r=\pm}\sum_{j=1}^{N_r}n_{r,j},
\end{eqnarray}
where $r=\pm$ labels the two different half-chains; $C_{r,j}^\dagger$ 
($C_{r,j}$)
are the creation (annihilation) operators of the spinless fermions; 
$n_{r,j}=C_{r,j}^\dagger C_{r,j}$
and $h$ here denotes the chemical potential. The Hamiltonian is equivalent 
to an $XXZ$ chain (\ref{hxxz}) via a Jordan-Wigner transformation.

We add a tunneling term to the Hamiltonian (\ref{spinless})
\begin{eqnarray}
{\bf T}=-V \left[C_{+,1}^\dagger C_{-,1}+C_{-,1}^\dagger C_{+,1} 
\right] \; , {~~~~~}V \ll 1 \; .
\end{eqnarray}
The tunneling current is thus
\begin{eqnarray}
J=-iV \left[C_{+,1}^\dagger C_{-,1}-C_{-,1}^\dagger C_{+,1} \right] \; .
\end{eqnarray}
From linear response theory we know that the tunneling conductance up to
order $V^2$ is given by
\begin{eqnarray}
\label{conduct}
G(\omega)=i\int dte^{i\omega t}\int dt'\theta(t-t')\int dt''\theta(t-t'')
\langle [J(t),J(t'')] \rangle \; .
\end{eqnarray}
Since the average $ \langle \cdots \rangle $ 
is taken at ${\bf T}=0$, the current correlation
 function may be separated into
\begin{eqnarray}
\langle [J(t),J(t'')] \rangle \sim \prod_{r=\pm} \langle C_{r,1}(t)
C_{r,1}^\dagger(t'') \rangle \sim (t-t'')^{-2\eta_\perp} \; ,
\end{eqnarray}
where $\eta_\perp$ is the boundary critical exponent of the single particle 
correlation function. Substituting the above relation into (\ref{conduct}) 
we readily obtain
\begin{eqnarray}
G(\omega) \sim \omega^\theta, {~~~~} \theta=2(\eta_\perp-1) = 
\frac{1}{Z^2(\Lambda)} - 2 \; .
\end{eqnarray}
At finite but very low temperatures $T \sim 0$, the conductance behaves as
\begin{eqnarray}
G(T)=G_0T^\theta,
\end{eqnarray}
where $G_0$ is a constant.
\par
At $U=0$, $H$ describes free fermions with a barrier. In this case, 
$\eta_\perp=1$ and $G(T)$ is independent of temperature and finite. 
The system is marginal. For $U > 0$, $\eta_\perp >1$ and the 
conductance tends to
zero as $T \to 0$.  The fermion-barrier scattering is relevant
and the ``Coulomb blockade'' behavior arises -- a result consistent with
the observations of Kane and Fisher \cite{13}. For $U<0$, $\eta_\perp < 1$ and 
the tunneling conductance diverges
as $T \to 0$. This is a consequence of the divergent superconducting 
fluctuations found in that situation. 
Our calculation may be generalized to general Luttinger liquids
and the critical exponent $\theta$ is then given by
\begin{eqnarray}
\theta=\sum_\nu K_\nu^{-1}-2,
\end{eqnarray}
where $K_\nu$ are the stiffness constants of the Luttinger liquid under 
consideration. 

\section{multicomponent integrable models}
Recently much attention has been focused on the open boundary problem for 
integrable
models with multi-component fields \cite{23,24,25}. 
Typical models are the one-dimensional $\delta$-potential Fermi gas model 
\cite{18}, the Hubbard 
chain \cite{22} and the supersymmetric $t-J$ model with open boundaries 
\cite{23,24}. 
The above discussion can also be generalized to these  models. In these 
cases, the reflection 
Bethe ansatz equations take the general form
\begin{eqnarray}
2Lp_0^\alpha(\lambda_j^\alpha)=2\pi I_j^\alpha-\sum_{\beta=1}^M
\sum_{l=1}^{N_\beta}{'}
[\Phi_{\alpha\beta}(\lambda_j^\alpha-\lambda_l^\beta)+
\Phi_{\alpha\beta}(\lambda_j^\alpha+\lambda_l^\beta)],
\end{eqnarray}
where $\Phi_{\alpha\beta}(\lambda_j^\alpha-\lambda_l^\beta)$ are the 
bare scattering phases and odd functions of their arguments, $M$ is the 
number of the components, and the prime after the sums
means that when $\alpha=\beta$, $j\neq l$. The eigenvalue
of the Hamiltonian is
\begin{eqnarray}
E=\sum_{\alpha=1}^M\sum_{j=1}^{N_\alpha}\epsilon_0^\alpha(\lambda_j^\alpha).
\end{eqnarray}
Also, $\lambda_j^\alpha > 0$ is supposed. In complete analogy to 
Eqs.~(\ref{rhob})--(\ref{drmomcon}) and (\ref{dressed}), dressed 
quantities are defined as
\begin{eqnarray}
\rho_b^\alpha(\lambda^\alpha)=\frac{p_0^\alpha{'}(\lambda^\alpha)}\pi
+\sum_{\beta=1}^M
\int_0^{\Lambda_\beta}K_{\alpha\beta}(\lambda^\alpha,\lambda^\beta)
\rho_b^\beta(\lambda^\beta)
d\lambda^\beta \;  , \\
p_b^\alpha(\lambda^\alpha)=p_0^\alpha(\lambda^\alpha)+
\frac12\sum_{\beta=1}^M\int_0^{\Lambda_\beta}
\left[\Phi_{\alpha\beta}(\lambda^\alpha-\lambda^\beta)+
\Phi_{\alpha\beta}(\lambda^\alpha+\lambda^\beta) \right] 
\rho_b^\beta(\lambda^\beta)d\lambda^\beta \; , \\
\epsilon_b^\alpha(\lambda^\alpha)=\epsilon_0^\alpha(\lambda^\alpha)+
\sum_{\beta=1}^M\int_0^{\Lambda_\beta}K_{\alpha\beta}(\lambda^\alpha,
\lambda^\beta)
\epsilon_b^\beta(\lambda^\beta)d\lambda^\beta \; , \\
Z_{\alpha\beta}^b(\lambda^\beta)=\delta_{\alpha\beta}+\sum_{\gamma=1}^M
\int_0^{\Lambda_\gamma}
Z_{\alpha\gamma}^b(\lambda^\gamma)K_{\gamma\beta}(\lambda^\gamma,
\lambda^\beta)d\lambda^\gamma 
\; ,
\end{eqnarray}
where $K_{\alpha\beta}(\lambda^\alpha,\lambda^\beta)=
\frac1{2\pi}[\Phi_{\alpha\beta}{'}(\lambda^\alpha-\lambda^\beta)+
\Phi_{\alpha\beta}{'}(\lambda^\alpha+\lambda^\beta)]$ is an even function. 
The relations 
\begin{eqnarray}
\frac\partial{\partial\lambda^\alpha}p_b^\alpha(\lambda^\alpha)=
\pi\rho_b^\alpha(\lambda^\alpha)
=2\pi\rho_\alpha(\lambda^\alpha) \; , \\
\epsilon_b^\alpha(\lambda^\alpha)=\epsilon_\alpha(\lambda^\alpha) \; , \\
Z_{\alpha\beta}^b(\lambda^\beta)=Z_{\alpha\beta}(\lambda^\beta) \; ,
\end{eqnarray}
for $\lambda^\alpha >0$, compare bounded and periodic systems at equal 
density and generalize 
(\ref{compare}). The finite size correction to the energies of the 
excited states is then
\begin{eqnarray}
E_{b,L}-E_{b,L}^0=\frac\pi L\sum_{\alpha=1}^Mv_\alpha 
\left\{\frac12[({\bf Z^{-1}\Delta N})_\alpha]^2+\Delta I_\alpha \right\}+
o(\frac1L),
\end{eqnarray}
where ${Z}_{\alpha\beta}={Z}_{\alpha\beta}(\Lambda_\beta)$, 
${\bf \Delta N}=\{\Delta N_1,...,\Delta N_M\}$ are $M$-dimensional vectors
with integer components. The number $\Delta N_\alpha$ gives the 
change of $N_\alpha$, the number of 
pseudoparticles of type $\alpha$ (pseudoparticles refers to the particle-like 
excitations in the Bethe ansatz and not necessarily to physical particles),
in the excited state with respect to the ground state (i.e. the charge
excitations). The non-negative integers $\Delta I_\alpha$ describe
pseudoparticle-pseudohole excitations 
(more precisely, a change of $\sum_{j=1}^{N_\alpha} 
p_b^\alpha(\lambda_j^\alpha)$ in units of
$\frac \pi L$) in the vicinity $k_F^\alpha$ (Fermi momenta $k_F^\alpha$ 
of the pseudoparticles are defined as $k_F^\alpha=
p_b^\alpha(\Lambda_\alpha)=\pi n_\alpha$). The Fermi velocity is
$v_\alpha=\frac{\epsilon_\alpha{'}(\Lambda_\alpha)}{2\pi\rho_\alpha
(\Lambda_\alpha)}$ as usual.
\par
The finite size correction of the ground state energy is given by
\begin{eqnarray}
E_{b,L}^0=\frac L\pi\sum_{\alpha=1}^M\int_0^{\Lambda_\alpha}p_0^\alpha{'}
(\lambda^\alpha)
\epsilon_b^\alpha(\lambda^\alpha)d\lambda^\alpha+f_b
-\frac\pi{24L}\sum_{\alpha=1}^Mv_\alpha+o(\frac 1{L}).
\end{eqnarray}
The Fermi velocities $v_\alpha$ are arbitrary in principle and 
quantitatively depend on details 
of the interactions in practice. As a consequence, the system is 
described by a sum of $M$
conformal algebras, each with a central charge 1. Their contributions to 
the boundary critical 
exponents 
\begin{eqnarray}
\label{boundalpha}
2 x_b^\alpha=[({\bf Z^{-1}\Delta N})_\alpha]^2+ 2 \Delta I_\alpha,
\end{eqnarray}
are additive, and the total boundary critical exponent is thus
\begin{eqnarray}
\label{boundsum}
2 x_b=2 \sum_{\alpha=1}^Mx_b^\alpha=2 \sum_{\alpha=1}^M\Delta I_\alpha
+ ({\bf Z^{-1}\Delta N})^T({\bf Z^{-1}\Delta N}).
\end{eqnarray}
The same structure is found in periodic systems.

As an example, we give some leading boundary critical exponents
of the Hubbard chain with open boundaries. The bulk critical exponents of 
this model were 
determined by Frahm and Korepin \cite{19}. The Hamiltonian reads
\begin{eqnarray}
H=-\sum_{i=1}^{N-1}\sum_{\sigma=\pm}C_{i\sigma}^\dagger C_{i+1\sigma}+
4U\sum_{i=1}^{N}n_{i\uparrow}n_{i\downarrow}
-\mu\sum_{i=1}^{N}\sum_{\sigma=\pm}n_{i\sigma}-
\frac h2\sum_{i=1}^{N}(n_{i\uparrow}-n_{i\downarrow}),
\end{eqnarray}
where $C_{i\sigma}$ ($C_{i\sigma}^\dagger$) is the electron annihilation 
(creation) operator;
$\mu$ denotes the chemical potential and $h$ is the external magnetic 
field. The wave functions 
are parametrized by two sets of parameters $\{k\}$ and  $\{\lambda\}$, 
the rapidities
of the charges and spins, respectively. The following set of integral 
equations determine
their bare ($\epsilon_0^{c,s}$) and dressed ($\epsilon_b^{c,s}$) 
energies and distribution 
functions ($\rho_b^{c,s}$)
\begin{eqnarray}
\epsilon_0^c(k)=-2\cos k+\mu-\frac h2,{~~~~~~~~}\epsilon_0^s(\lambda)=
\frac h2,\nonumber\\
\epsilon_b^c(k)=\epsilon_0^c(k)+\int_0^{\Lambda_s}K_1(\sin k,\lambda)
\epsilon_b^s(\lambda)d\lambda,\nonumber\\
\epsilon_b^s(\lambda)=\epsilon_0^s(\lambda)+
\int_0^{\Lambda_c}\cos kK_1(\lambda,\sin k)\epsilon_b^c(k)dk
-\int_0^{\Lambda_s}K_2(\lambda,\mu)\epsilon_b^s(\mu)d\mu,\\
\rho_b^c(k)=\frac 1\pi+\cos k\int_0^{\Lambda_s}K_1(\sin k,\lambda)
\rho_b^s(\lambda)d\lambda,\nonumber\\
\rho_b^s(\lambda)=\int_0^{\Lambda_c}K_1(\lambda,\sin k)\rho_b^c(k)dk-
\int_0^{\Lambda_s}K_2(\lambda,\mu)\rho_b^s(\mu)d\mu.
\end{eqnarray}
The dressed charge (\ref{dressed}), in the multicomponent problem, takes a 
matrix structure, with elements 
\begin{eqnarray}
Z_{cc}^b(k)=1+\int_0^{\Lambda_s}Z_{cs}^b(\lambda)
K_1(\lambda,\sin k)d\lambda,\nonumber\\
Z_{cs}^b(\lambda)=\int_0^{\Lambda_c}\cos kZ_{cc}^b(k)K_1(\sin k,\lambda)
dk-\int_0^{\Lambda_s}Z_{cs}^b(\mu)K_2(\mu,\lambda)d\mu,\nonumber\\
\label{dressmat}
Z_{sc}^b(k)=\int_0^{\Lambda_s}Z_{ss}^b(\lambda)K_1(\lambda,\sin k)d\lambda,\\
Z_{ss}^b(\lambda)=1+\int_0^{\Lambda_c}\cos kZ_{sc}(k)K_1(\sin k,\lambda)dk-
\int_0^{\Lambda_s}Z_{ss}^b(\mu)K_2(\mu,\lambda)d\mu,\nonumber
\end{eqnarray}
with the kernel
\begin{eqnarray}
K_n(\lambda,\mu)=\frac 1\pi[\frac{nU}{(nU)^2+(\lambda-\mu)^2}
+\frac{nU}{(nU)^2+(\lambda+\mu)^2}], {~~~~~~~}n=1,2.
\end{eqnarray}
As in the single-component case, the dressed charge matrices for open and
periodic boundary conditions are identical, 
$Z_{\alpha,\beta}^b = Z_{\alpha,\beta}$. The Fermi velocities are given by
\begin{eqnarray}
v_c=\frac{\epsilon_b^c{'}(\Lambda_c)}{\pi\rho_b^c(\Lambda_c)},
{~~~~~~~~}v_s=\frac{\epsilon_b^s{'}(\Lambda_s)}{\pi\rho_b^s(\Lambda_s)},
\end{eqnarray}
and $\Lambda_{c,s}$ are defined by $\epsilon_b^{c,s}(\Lambda_{c,s})=0$.
\par
Below we list some correlation functions  and the $\Delta N_{c,s}$, 
$\Delta I_{c,s}$ which
must be used in Eqs.~(\ref{boundalpha}) and (\ref{boundsum}) in order 
to determine
the leading critical exponents:\\
(1) The field correlator
\begin{eqnarray}
G_{\Psi\Psi}(x_1,x_2,t)= \langle C_{x_1\uparrow}(t)C_{x_2\uparrow}^\dagger(0)
\rangle ,\\
\Delta N_c=1,{~~~}\Delta N_s=0, {~~~}\Delta I_c=\Delta I_s=0.\nonumber
\end{eqnarray}
(2) The density-density correlator
\begin{eqnarray}
G_{nn}(x_1,x_2,t)= \langle n_{x_1}(t)n_{x_2}(0) \rangle \; ,\\
\Delta N_c=\Delta N_s=0, {~~~}\Delta I_c=1,{~~~} \Delta I_s=0 {~~~}{\rm or}
{~~~} \Delta I_c=0,{~~~} \Delta I_s=1.\nonumber
\end{eqnarray}
(3) The spin-spin correlators
\begin{eqnarray}
G_{\sigma\sigma}^z(x_1,x_2,t)= \langle S^z(x_1,t)S^z(x_2,0) \rangle \; ,\\
S^z(x,t)=\frac12[n_{x\uparrow}-n_{x\downarrow}],\nonumber\\
\Delta N_c=\Delta N_s=0, {~~~}\Delta I_c=1,{~~~} \Delta I_s=0 {~~~}
{\rm or } {~~~} \Delta I_c=0,{~~~} \Delta I_s=1.\nonumber
\end{eqnarray}
\begin{eqnarray}
G_{\sigma\sigma}^\perp(x_1,x_2,t)= \langle S^-(x_1,t)S^+(x_2,0) \rangle \; ,\\
S^+(x,t)=C_{x\uparrow}^\dagger(t) C_{x\downarrow}(t),\nonumber\\
\Delta N_c=0,{~~~} \Delta N_s=1, {~~~}\Delta I_c=\Delta I_s=0.\nonumber
\end{eqnarray}
(4) The triplet pair correlator
\begin{eqnarray}
G_p^{(1)}(x_1,x_2,t)= \langle C_{x_1+1\uparrow}(t) 
C_{x_1\uparrow}(t) 
C_{x_2\uparrow}^\dagger(0)C_{x_2+1\uparrow}^\dagger(0) \rangle \; ,\\
\Delta N_c=2,{~~~} \Delta N_s=0, {~~~}\Delta I_c=\Delta I_s=0.\nonumber
\end{eqnarray}
(5) The singlet pair correlator
\begin{eqnarray}
G_p^{(0)}(x_1,x_2,t)=\langle C_{x_1\uparrow}(t)
C_{x_1\downarrow}(t)
C_{x_2\downarrow}^\dagger(0)C_{x_2\uparrow}^\dagger(0) \rangle \; ,\\
\Delta N_c=2,{~~~} \Delta N_s=1, {~~~}\Delta I_c=\Delta I_s=0.\nonumber
\end{eqnarray}
Precise values for the critical exponents then follow immediately, via 
(\ref{boundalpha})
and (\ref{boundsum}), once the dressed charge matrix (\ref{dressmat}) 
is calculated. This
is a matter of routine, and due to the equality of this matrix
for open and periodic systems (cf.~above), the published results for the
periodic Hubbard model \cite{19}
can be used directly to evaluate the boundary critical expoenents.

\section{summary}
We have derived explicitly the boundary critical exponents of both 
single- and multi-component
Bethe ansatz soluble models of interacting bosons and fermions. 
Our results imply that the descendant fields (particle-hole excitations) 
contribute the same
(integer) amount to the boundary and the bulk  critical exponents. 
However, the contribution
from charge excitations (additional
particles) to the boundary critical exponents is twice as big as to 
the bulk exponents. The current excitations are completely 
depressed for open boundaries and 
thus contribute nothing
to the boundary critical exponents. Apparently, this statement 
is valid much beyond the the Bethe
ansatz soluble models and applies in general to Luttinger liquids with
open boundaries \cite{12}. The critical exponents are determined by 
the dressed charge matrix 
which we have shown to be independent of the boundary conditions. 
Moreover, our method of calculation
relies only on the determination of energies which can be performed 
accurately by numerical
methods in models which cannot be solved by Bethe ansatz. Therefore, 
one can determine, at
least numerically, the boundary critical exponents for all 1D 
quantum systems, provided they
are conformally invariant, by the method described in this paper.

\acknowledgements

Two of the authors acknowledge support from the Max-Plank-Gesellschaft and the 
Alexander von Humboldt-Stiftung (Y. W.) and from 
Deutsche Forschungsgemeinschaft under SFB 279-B4 (J.V.).

\end{document}